# The spin-off of elite universities in non-competitive, undifferentiated higher education systems: an empirical simulation in Italy[1]

## Abstract

Higher education systems featuring intense competition have developed world-class universities, capable of attracting top professors and students and considerable public-private funding. This does not occur in non-competitive systems, where highly-talented faculty and students are dispersed across all institutions. In such systems, the authors propose the budding of spin-off universities, staffed by migration of top scientists from the entire public research system. This work illustrate the proposal through an example: the spin-off of a new university in Rome-Italy staffed with the best professors from the three current public city universities. Such a faculty would offer top national research productivity, a magnet to attract the other critical ingredients of a world-class university: talented students, abundant resources and visionary governance.

---

## 1. Introduction

In recent decades, higher education systems throughout the world have experienced strong expansion in both demand and offer. There are constant increases in the number of subjects operating in the global system, in their geographic distribution, and in the extent of diversification at the level of disciplinary specialization (Teixeira et al., 2011). These changes respond to the growing "massification" of higher education (Rossi, 2010) but also result from interventions by policy makers to reinforce competitive market-like mechanisms and thus achieve greater effectiveness and efficiency (Jongbloed, 2004). The processes of globalization impact on higher education just as much as on the other sectors. For some years, world-class universities that have traditionally attracted highly talented foreign faculty, students and financing have also internationalized their activity by opening satellite campuses in other nations. Given the current knowledge-based economy, governments are more often making decided efforts to improve the average quality of their higher education systems and to stimulate the growth of elite universities. The European Union considers the contribution of higher education as central in giving the region "the most competitive economy and knowledge-based society of the 21st century". There is awareness that the European higher education system is still too fragmentary and hampered by a combination of excessive public control and scarce autonomy. This has led many national governments to begin reforms aimed at releasing the full potential of their domestic universities. In this context, the particular social and economic benefits of elite universities to the nation and their home regions have been soundly demonstrated (Yusuf and Nabeshima, 2007; Fritsch and Slavtchev, 2007; Rosenberg and Nelson, 1994; Jaffe, 1989). While non-elite universities also play a substantial role in fostering national and regional development, the impact of elite universities is unquestionable, even though the extent of their influence depends in part on the characteristics of the national and regional economic systems and normative environments.

The aim of the current work is to contribute to a more evidence-based reform process by providing empirical simulation of the creation of elite[2] universities in non-competitive and undifferentiated higher education systems.

The first step in designing policies to foster elite universities is to define what they have that others lack and to identify the catalysts for the development of such characteristics. Several scholars have identified the distinctive features of elite universities, indicating the presence of highly qualified faculty, talented students, abundant resources, autonomy, and favorable governance (Niland, 2007; Altbach, 2004). The growth of elite universities is favored by regulatory environments that introduce and permit competitive mechanisms for the stimulation of continuous improvement and the pursuit of competitive advantage. Aghion et al. (2009) show that universities' performance is correlated with their autonomy and competitive environment. The level of competitiveness depends on cultural and other contextual factors, particularly the level of university autonomy, the type of financing and a government regulatory framework that is supportive in character. In competitive higher education systems, such as those observed in English speaking nations, the pursuit of competitive advantage has led to development of world-class universities that can attract, develop and retain highly-talented national and foreign faculty and students. The same institutions also obtain abundant public financing, private financing and donations, and attract venture capital and establishments of national and international enterprise in their territory, with resulting social and economic benefits. The competition factor has generated universities that are distinct in their quality of education and research, and thus in prestige, and that offer degrees and produce research results that also stand out for their social and market value. However, in nations where the state has a significant presence in the regulation and direct financing of universities (Auranen and Nieminen, 2010) and where competitive aspects of higher education are scarce or absent, we observe the contrary: here we see little differentiation among universities and an absence of elite institutions. While we accept the limits of international comparisons (Abramo et al., 2011a; Turner, 2005), it is no accident that the 2011 SJTU Academic Ranking of World Universities[3], lists the entire top 20 institutions as being U.S. (with 17) or U.K. (three), while the THES 2011-2012 World Academic Ranking[4] includes only one Swiss and one Canadian university among the top 20, which are otherwise again all from the U.K. and the U.S.

[2] We distinguish between elite and world-class universities: a university that is elite in a particular country is not necessarily world-class. The term "elite universities" is referring to institutions displaying top-level performance in research.

Without doubt, higher education is continually assuming the character of a global market with local impact, where competition among universities is ever more international. Countries without elite universities inevitably undergo brain drain of highly talented faculty and students towards nations with world-class institutions. There is also a drain of financial resources, as private companies prefer to finance research projects in foreign world-class universities rather than domestic ones. If "have not" countries wish to gain prominent institutions that compete at the global level, then the role of the state becomes fundamental. In the past, such universities as Oxford, Cambridge or the U.S. Ivy League would grow to prominence as a result of incremental progress, rather than by deliberate government intervention. It is impossible to think of rapidly creating elite universities at the present moment, without direct intervention and support from the state (Salmi, 2009). Through survey and analysis of modern international experience, three possible strategies have been identified (Salmi, 2009): i) focus resources on a restricted number of existing universities with the greatest potential (a strategy recently adopted by the French government and highly recognizable in the "German Initiative for Excellence"); ii) encourage mergers of a number of existing universities to realize the necessary synergies for becoming world-class universities (the governments of Denmark, China and Russia have policy to reward the merger of similar institutions; Cardiff University and the South Wales School of Medicine have deliberately merged as a step towards a world-class university in Wales); iii) create new world-class universities from a fresh start (examples are mainly seen in emerging countries such as India, Kazakhstan and Saudi Arabia, but also in Europe, as in the Paris School of Economics).

The authors, stimulated by the availability of sophisticated bibliometric instruments and by their extensive observation of the Italian higher education system, which has very little differentiation and definitely lacks elite universities of meaningful size, suggest a fourth strategy: to bud spin-off universities from existing ones, through the extraction of top professors from the parent universities. A highly talented faculty is not alone sufficient to create an elite university, but in our opinion it is the single most important factor, and can catalyze attraction of all the other determinants (talented students, abundant resources, appropriate governance). Government must still play an essential role, in establishing a favorable regulatory environment and supporting the internationalization and global competitiveness of spin-off universities during the start-up phase.

[3] http://www.shanghairanking.com/ARWU2011.html
[4] http://www.timeshighereducation.co.uk/world-university-rankings/2011-2012/top-400.html

While we propose "budding" as a strategy for international consideration, the characteristics of the Italian academic system permit a ready test of its feasibility. This is a system where universities have only recently achieved certain management autonomy, but where there are still tight restrictions and excessive central regulation, particularly affecting recruitment, salaries and awarding of degrees.

In this work we simulate the budding of a spin-off university from those already present in Rome, in order to avoid any obligation that professors move geographically, which could be a social deterrent that would inhibit creation of the new elite university. The procedures could be repeated for other areas of Italy, with the objective of spinning off a fairly homogenous national distribution of elite institutions, always beginning from existing universities. In our simulation the new Rome Spin-off University (RSU) would consist of five schools: mathematics, physics, chemistry, earth science and engineering. Of the 15 universities in Rome, only the three largest ones offer undergraduate degrees in these disciplines: the University of Rome "Sapienza", University of Rome "Tor Vergata" and University of Rome "Tre"[5]. According to the size of the teaching staff, these institutions rank as largest, 11$^{th}$ and 25$^{th}$ out of the 95 Italian universities. The spin-off professors are extracted from the three parent universities on the basis of scientific merit[6]. This is assessed with bibliometric techniques, thoroughly described and tested in literature that measure the research productivity of each professor and provide comparisons to all national colleagues of the same research field and academic rank (Abramo and D'Angelo, 2011).

In Sections 2 and 3 of this paper we summarize the characteristics of the Italian higher education system and illustrate the specific techniques for the dimensioning of the new RSU. In Section 4 we present the methodological approach for identifying the RSU professors and show the faculty composition. Section 5 presents statistics describing the distribution of individual performance by the selected professors compared to their national colleagues, and the performance of the overall RSU, at the level of fields and disciplines. The work concludes with a summary of the proposal and the authors comments on policy implications and implementation.

## 2. The Italian higher education system

The Italian Ministry of Education, Universities and Research (MIUR) officially recognizes a total of 95 universities, giving them authority to issue legally-recognized degrees. Twenty-eight of these are very small private special-focus universities, of which 12 offer only e-learning. Sixty-seven are public and generally multi-disciplinary universities, scattered throughout the nation, some having a number of branches in smaller towns. Six institutions are *Scuole Superiori* (Schools for Advanced Studies), specifically devoted to highly talented students, with very small faculties and tightly limited enrollment numbers per degree program. In Italy, 94.9% of faculty are employed in public universities (0.5% in Scuole Superiori) and 5.1% are in private universities.

[5] This university does not have a school of chemistry.
[6] We assume that quality in teaching will also follow from scientific merit.

The Italian higher education system is a long-standing classic example of a public and highly centralized governance structure, with low levels of autonomy at the university level and a very strong role played by the central state. Until 2008, core government funding was input oriented, and distributed to satisfy the resource needs of each and every university in function of their size and activities. It has only been since 2009, following the first national research assessment exercise (VTR), that a minimal share of MIUR financing has been allocated based on research and teaching quality: this share represents 3.9% of total university income.

In keeping with the German "Humboldt" model, all university professors are contractually obligated to carry out research, thus there are no teaching-only institutions in Italy. National regulations establish that each faculty member must allocate a minimum of 350 hours per year to teaching. At the end of 2010, there were 58,000 faculty members in Italy (full, associate and assistant professors) and a roughly equal number of technical-administrative staff. All new personnel enter the university system through public examinations and career advancement can only proceed by further public examinations. Salaries are regulated at the centralized level and are calculated according to role (administrative, technical, or professorial), rank within role (for example: assistant, associate or full professor) and seniority. None of a professor's salary depends on merit: salaries increase annually according to rules set by government. Moreover, as it is throughout the Italian public administration, dismissal of an employee for lack of productivity is unheard of.

The whole of these conditions create an environment and a culture that are completely non-competitive, yet flourishing with favoritism and other opportunistic behaviors that are dysfunctional to the social and economic roles of the higher education system. The overall result is a system of universities that are almost completely undifferentiated according to quality and prestige, with the exception of the tiny Scuole Superiori and a very small number of the private special-focus universities. The system is thus unable to attract significant foreign faculty or students. The numbers are negligible: foreign students are 3% of the total, compared to the OECD average of 8.5%, and only 2.3% of actual graduates are foreigners; only 1.8% of faculty are foreign nationals. This is a system where every university has some share of top professors, flanked by another share of absolute non-producers, an observation confirmed by empirical evaluation of the scientific performance of universities and individual faculty members (Abramo et al., 2011b). Over the 2004-2008 period, 6,640 (16.8%) of the 39,512 hard sciences professors did not publish any scientific articles in the journals censused by the Thomson-Reuters *Web of Science* (WoS). Another 3,070 professors (7.8%) did achieve publication, but their work was never cited. This means that 9,710 individuals (24.6%) had no impact on scientific progress. An almost equal 23% of professors alone produced 77% of the overall scientific advancement (Abramo et al., 2011b). The problem is that most productive faculty are not concentrated in a limited number of universities, but is instead dispersed more or less uniformly among all Italian universities, along with the unproductive individuals, so that no single institution reaches the critical mass of excellence necessary to develop as an elite university and compete at the international level. Abramo et al. (2012a) show empirical evidence that research performance across Italian universities has a more or less similar pattern of distribution. The authors found that the performance distribution within universities is generally highly concentrated on certain individuals, similar to the pattern of the entire research population. They also found that the variability of average performance among universities is below that for the subpopulation of individuals within each university.

The Italian government has granted universities very much greater autonomy, but has done very little to create disciplined competition for research funding, faculty, and students. Autonomy and competition together would have important synergistic effects, but one of these alone can be very dangerous. There is a widely-shared opinion that institutional evaluation with subsequent selective allocation of funding could resolve a good part of the obvious inefficiencies, but the authors remain skeptical. In cases such as Italy's, where accompanying salary schemes based on merit are missing and where the share of resource allocation based on institutional merit is very small, it is doubtful that such a system would lead to significant increments in production efficiency, or displace historic practices of favoritism and give way to efficient selection processes. Given this, it is still more difficult to believe that we will see the development of elite universities, even over the long term. Given the knowledge-based economy and the urgency of global competitive challenges, the authors suggest a more daring policy: the budding of spin-off universities staffed by migration from only the top scientists of the national research system, possibly with the new institutions located in a balanced regional manner. With very low investment, it would be possible to quickly create the type of elite universities that other competitive systems have produced over decades of time: universities capable of attracting the best faculty and students and public and private capital, and of providing much greater economic benefit than the present universities, with their high dispersion of internal performance.

## 3. Determining the size of Rome Spin-off University

The three largest universities in Rome are generalist in character. Although they show some diversification, all three are active in the hard sciences, social sciences and humanities. The oldest and largest of the three universities, Rome "Sapienza", is situated in the city center and has a faculty of over 4000. The two newer institutions are the University of Rome "Tor Vergata", founded in 1982 and University of Rome "Tre", founded in 1992, both based further from the city center and both established in order to relieve crowding at the Sapienza, which had reached the limits of governability and was no longer capable of meeting the increasing demand for education.

In the Italian university system each professor is classified in one and only one research field. There are a total of 370 such fields (named scientific disciplinary sectors, or SDSs[7]), grouped into 14 disciplines (named university disciplinary areas, or UDAs). The dimension and scope of the research fields that we propose for the RSU depend, on the one hand, on the choice to use bibliometrics as a starting point for preparing the empirical simulation, and, on the other hand, on the choice of parent institutions. Bibliometric methodology is applicable only to the hard sciences and restricts our proposal for the RSU faculty only to those specializations. The faculty size and field specialization of the RSU are based on those of Rome Tre, the smallest of the three generalist institutions, thus guaranteeing true feasibility in implementing the RSU. For the same purpose, we also try as much as possible to reproduce the distribution of fields, disciplines and academic ranks of Rome "Tre". This is possible for four of the five RSU schools: mathematics, physics, earth sciences and engineering. Rome "Tre" currently does not have a separate department of chemistry so we take the faculty of the department of chemistry of Rome "Sapienza" and rescale its size according to the faculty ratio of the two institutions for the other considered UDAs [8]. The RSU should thus have 247 professors, with distribution per SDS and UDA as presented in Table 1. The sizing criteria should permit the RSU to establish at least eight degree programs: Mathematics, Physics, Chemistry, Earth sciences, Civil engineering, Mechanical engineering, Electronics engineering and Computer science.

[7] The complete list is accessible at http://attiministeriali.miur.it/UserFiles/115.htm. Last accessed on January 21, 2013.

The RSU would be medium-sized for overall numbers of faculty, placing 46th out of the total 90 Italian universities. We could "enlarge" the empirical simulation by broadening the range of parent institutions to universities in other regions or to other public research organizations, thus permitting an increase in the number of faculty and range of disciplines. The adoption of peer-review methodology (difficult to simulate for the current purposes) would also permit the inclusion of the social sciences, arts and humanities. However neither the size of faculty nor the scope of research fields would affect the key issue of research productivity since there are no returns to size (Abramo et al., 2012b; Bonaccorsi and Daraio, 2005) and also no returns to scope (Abramo et al., 2012c).

[Table 1]

## 4. Selecting RSU faculty

The basic idea of the model is that, beginning from the three generalist Rome universities, the RSU will draw the professors with the best scientific performance in the 64 SDSs indicated in Table 1. Since the university has a special focus on the hard sciences, the literature gives ample indication (Moed, 2005; Glänzel and Debackere, 2003) that it is legitimate to use bibliometric techniques to measure scientific performance. In the next section we give the details of the selection of RSU faculty.

### 4.1 Methodology and dataset

To assess scientific productivity of individual researchers we consider the outcome, or impact of their research activities, over the five year period from 2004 to 2008. As proxy of outcome we adopt the number of citations for the researcher's publications at 30/06/2009. Because the intensity of publications varies by field, we compare researchers within the same field, meaning the same SDS (Abramo and D'Angelo, 2011). When measuring labor productivity, if there are differences in the production factors available to each scientist then there should productivity should be normalized according to them. Unfortunately relevant data are not available in Italy. Another issue is that it is very possible that researchers belonging to a particular scientific field will also publish outside that field. Because citation behavior varies by field, we standardize the citations for each publication with respect to the median[9] of the distribution of citations for all the Italian cited-only publications of the same year and the same WoS subject category[10]. Furthermore, research projects frequently involve a team of researchers, which shows in co-authorship of publications. In this case we account for the fractional contributions of scientists to outputs, as the reciprocal of number of co-authors. The productivity of a single researcher ($P_R$)[11], is given by:

$$P_R = \sum_{i=1}^{n} \frac{c_i}{m_i} * \frac{1}{s_i} \qquad [1]$$

[8] Including other disciplines, such as medicine, in the RSU would have restricted the selection of the top-scientists for these fields to only two "parent" universities, which would have jeopardized the aim of the simulation, i.e. the realization of a university reaching elite status in all disciplines. For the same reason we have limited the RSU size to that of Rome "Tre" thus ensuring that, like the smallest of its parent universities, it would be capable of providing at least the same number of bachelor and masters programmes in all disciplines. We note that by expanding the range of parent universities beyond Rome, it would be possible to simulate the development of larger and more generalist spin-off universities.

Where:
$c_i$ = citations received by publication *i*;
$m_i$ = median of the distribution of citations received for all Italian cited-only publications of the same year and subject category of publication *i*;
$s_i$ = co-authors of publication *i*
n = number of publications of the researcher in the period of observation.

Data on faculty of each university and their SDS classification is extracted from the database on Italian university personnel, maintained by the MIUR[12]. The bibliometric dataset used to measure $P_R$ is extracted from the Italian Observatory of Public Research (ORP)[13], a database developed and maintained by the authors and derived under license from the Thomson Reuters WoS. Beginning from the raw data of the WoS, and applying a complex algorithm for reconciliation of the author's affiliation and disambiguation of the true identity of the authors, each publication (article, article review and conference proceeding) is attributed to the university scientist or scientists that produced it (D'Angelo et al., 2010). We then elaborate $P_R$ ranking lists for each SDS, for the publication window 2004-2008. For each SDS, we then extract the names of the best professors active in each of the three parent Rome universities from the rankings lists, in numbers as indicated in Table 1. The next section shows the results of this extraction.

## 4.2 Composition of the RSU faculty

Of the 247 professors selected, 87 (35.2% of total) are full professors, 88 (35.6%)

[9] As frequently observed in literature (Lundberg, 2007), standardization of citations with respect to median value rather than to the average is justified by the fact that distribution of citations is highly skewed in almost all disciplines.

[10] The subject category of a publication corresponds to that of the journal where it is published. For publications in multidisciplinary journals the standardized value is calculated as the average of standardized values for each subject category.

[11] This indicator is similar to the "crown indicator" of Leiden's CWTS (Moed et al., 1985) and the "total field normalized citation score" of the Karolinska Institute (Rehn et al., 2007). The differences are: i) we standardize citations of single publications and not of scientific portfolio of researchers; ii) we standardize by the Italian median rather than the world average. We also consider fractional counting of citations based on co-authorship.

[12] http://cercauniversita.cineca.it/php5/docenti/cerca.php. Last accessed on January 21, 2013.

[13] www.orp.researchvalue.it. Last accessed on January 21, 2013.

are associates and 72 (29.1%) are assistants. We check this balance among the three academic ranks to guarantee that it would permit the new organization to meet both research and teaching needs. To do this we compare the distribution per academic rank of the RSU faculty against the average national distribution (last line, Table 2), and observe that the differences are acceptable. The greatest variation (-3.8%), for assistant professors, is not enough to cause any particular worries.

[Table 2]

At the level of single UDAs, such variations clearly tend to differ, and in some cases could seem problematic: in physics, assistant professors are less than 20% of total faculty (8 out of 42); in mathematics they are 22% (6 of 27). However, if we analyze the same type of data for other individual Italian universities, we discover that such “unbalance” is not rare. For example, the International School for Advanced Studies in Trieste, which is one of the six Scuole Superiori, shows exactly the same type of distribution (46% full, 35% associate and 19% assistant professors). Thus in terms of the mix of academic ranks, the extraction process seems to have provided the RSU with a faculty composition that is appropriate for its institutional aims.

In terms of the contribution from the three parent Rome universities, 55% of the total RSU faculty comes from Sapienza, 26% from Tor Vergata and 19% from Tre. These results are shown Table 3, including the distribution per UDA.

[Table 3]

At the macro level, the data seem coherent with sizes of the parent universities. However an analysis of concentration by UDA offers an interesting view of the real contribution of faculty from each of the three parent universities to the form of the new RSU. Table 4 presents the analysis by UDA. The concentration index shown in parentheses[14] indicates that Physics is dominated by the professors originating from the Sapienza and Rome Tre. Rome Tre also contributes very significantly to the development of RSU faculty in Mathematics, Earth sciences and Engineering. The last line of Table 4 actually shows that, proportionally, Rome Tre contributes more to the formation of RSU faculty than the other two universities.

[Table 4]

## 5. RSU faculty performance

In this section we compare research performance between the RSU and all other Italian universities. We begin with a comparison at the level of individual professors. Then aggregating the data by SDS, we compare performance at the level of SDSs, UDAs, and finally the overall universities.

[14] An index value greater than one indicates that the incidence of faculty from the university considered is greater than expected given the total size of its faculty in that UDA.

### 5.1 RSU research performance at the individual level

The performance ranking lists are expressed in percentiles and differentiated by academic rank, since we have demonstrated that productivity of full, associate and assistant professors is different (Abramo et al. 2011c). Thus the performance of each professor is calculated in each SDS for each academic rank and expressed on a scale of 0-100 (worst to best) for comparison with the performance ($P_R$) of all Italian colleagues of the same academic rank and SDS.

Given the means of selection, we expect the bibliometric performance values for RSU faculty members to be very high: in this section we check these expectations. Table 5 shows the research performance of RSU professors by UDA, in terms of: i) average percentile of performance ($P_R$); ii) percentage of those with performance rank above $80^{th}$ national percentile; iii) percentage of those with performance rank above $90^{th}$ percentile. For purposes of comparison, we present the same data for the three parent universities and the top universities in each of three national subgroups: public, Scuola Superiore and private universities. The top public university is close to the same size as the RSU in the disciplines considered, while the top private university and Scuola Superiore are very small even considering the entire faculty. The data show unmistakably high performance: the average percentile value for RSU faculty is 88.8, with a peak of 95.0 in Chemistry, followed by Engineering (92.9) and Physics (91.2). To have a further idea of the level of concentration of excellence achieved by the new RSU, we compare to the average values registered by the parent universities and top Italian universities. The overall average performance of the three parent university faculties varies between 40.9 for Sapienza and 48.8 for Tor Vergata, and the best showing for other Italian universities is 69.6. In the RSU, 84.2% of the total faculty show a performance superior to the $80^{th}$ national percentile. In Chemistry, all professors selected for the RSU exceed the $80^{th}$ national percentile and for Engineering the percentage is 91.3. Meanwhile, in the three parent universities the level of "top 20%" professors varies between 14.9% and 21.8% (Sapienza and Roma Tre). In the best public Italian university this figure is 36.5% and it is 54.5% in the top Scuola Superiore.

If we raise the threshold to the $90^{th}$ percentile, the absolute excellence of the RSU faculty is clear: 57.5% of the 247 professors selected show a $P_R$ that puts them above the 90% threshold in comparison with all other Italian colleagues in the same SDS and academic rank, with a peak of 77.89% in Chemistry. In contrast, in the three parent universities the share of professors with performances above the $90^{th}$ percentile varies between 7.1% and 10.9%; the best Italian public university registers 22.8% and the best Scuola Superiore achieves 31.8%.

[Table 5]

Table 6 shows detailed information concerning the distribution of performance values for the 247 selected professors: for 17 of these (6.9% of total) there are no Italian colleagues of the same SDS and academic rank with higher $P_R$. In the three Rome parent universities, this situation occurs for only 0.8% of the total staff, and for Italian universities as a whole, this share is 1.1%. The concentration of top national professors in the RSU is thus eight times the average of the parent universities and more than six times the Italian average. Also, 9.7% of the selected professors register a performance value greater than or equal to the $99^{th}$ national percentile, compared to just 1.3% of professors in the parent university faculties and 1.6% of total Italian university faculty. Again, 38.1% of RSU professors, or seven times the average for Italian universities, place in the top 5% of professors in Italy. Only 15.8% of selected professors (39 out of 247), classify under the 80th percentile. Finally, only 1.2% (3) professors show a $P_R$ value below the national median (last line of Table 6), compared to 50.4% of total faculty in the parent universities.

[Table 6]

### 5.2 RSU research performance at the SDS level

We now analyze RSU performance at the scientific field level, meaning the SDS. A university SDS usually consists of professors of different academic ranks, with salaries dependent on their academic ranks and seniority. The impact of the SDS is given by the standardized citations for the publications of its member professors. Productivity $P_S$ is given by the overall impact divided by the number of professors, each weighted for their cost equivalent. The cost equivalent *C* shown in the last column of Table 7 is derived from MIUR[15] data on the average yearly cost of professors in each academic rank. In formula:

$$P_S = \frac{\sum_{i=1}^{n} P_{R_i}}{\sum_{i=1}^{n} C_i} \qquad [2]$$

where:
$P_{Ri}$ = productivity of the *i-th* professor in SDS
$C_i$ = cost equivalent of the *i-th* professor
$n$ = number of professors in SDS

[Table 7]

Calculating the productivity values as in [2] for all Italian universities active in the same SDS, we can then obtain the performance rank (absolute and percentile) for each of the RSU's SDSs (Table 8).

Of its 64 SDSs, there are 52 cases (81% of total) where the RSU's bibliometric performance is above the 80th national percentile. Almost 72% of SDSs (46 of 64) show productivity performance that is equal to or greater than the 90th national percentile. There are 21 (33%) SDSs where the RSU would lead the national rankings and another eight (12.5%) where it would place second. There are only two SDSs (ING-INF/07 and ING-INF/06) where RSU performance is equal to or below the national median.

[Table 8]

### 5.3 RSU research performance at UDA level

[15] https://dalia.cineca.it/php4/inizio_access_cnvsu.php. Last accessed on January 21, 2013

A UDA is formed of all professors belonging to its component SDSs. Performance evaluation for the UDA is conducted through standardization and weighting of the performance recorded at SDS level. Standardization to the national mean of the performance distribution for all universities active in the same SDS permits elimination of bias due to the different fertility of the SDSs; weighting with respect to the cost share of the SDS towards the total for the UDA takes into account the varying representativity of the SDSs in each UDA, in terms of faculty numbers and rank. This permits robust ratings in spite of the intrinsic heterogeneity of the SDSs (Abramo et al., 2008).

Productivity ($P_u$) of a generic UDA is calculated:

$$P_u = \sum_{i=1}^{n_u} \frac{P_i}{\overline{P_i}} \cdot \frac{C_i}{C_u} \qquad [3]$$

where:

$P_i$ = productivity of SDS $i$,

$\overline{P_i}$ = average value of productivity for Italian universities in SDS $i$

$C_i$ = cost equivalent of SDS $i$,

$C_u$ = cost equivalent of UDA $u$,

$n_u$ = number of SDSs active in UDA $u$.

We calculate the productivity values for each UDA at the RSU and for all Italian universities active in the same UDA. Results from the comparisons are shown in Table 9.

[Table 9]

Where it doesn't place first, the RSU is only outperformed by one or two of the tiny Scuole Superiori, however none of these has more than three professors in the UDAs examined.

Extending the summation [3] to all SDSs at the RSU, we arrive at the calculation of the university's overall productivity: the spin-off university achieves a performance value that would place it first in the national rankings among the 83 universities active in the hard sciences (Figure 1). The RSU's performance would exceed the national median by 200%, and that of the top public institution by 75%.

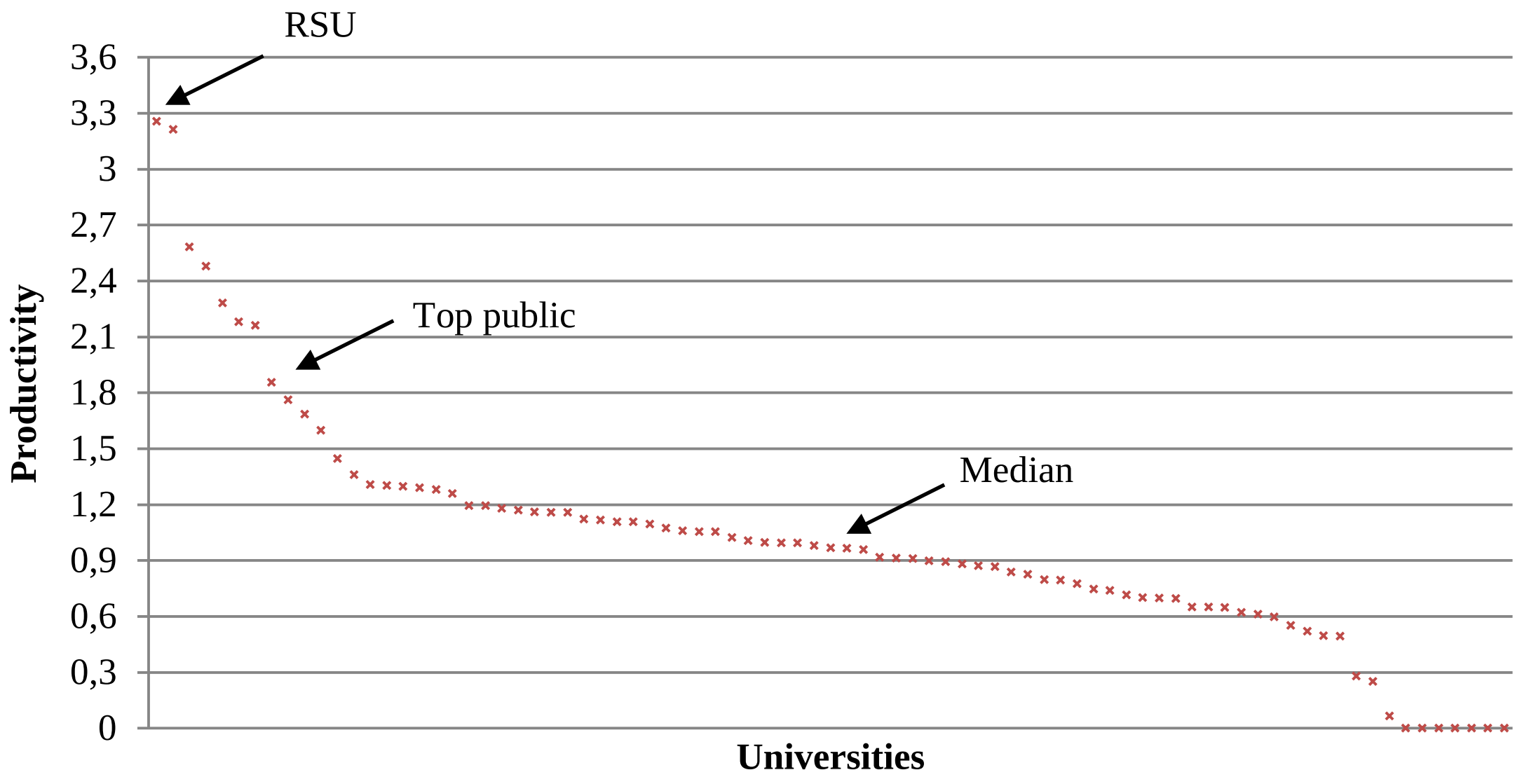

***Figure 1: Productivity distribution of Italian universities active in the hard sciences***

## 6. Discussion and conclusions

The challenges of global competition in research and higher education demand courageous and far-sighted policy choices. Some nations are already prepared to it because their systems have strong aspects of competition, particularly for resource allocation and in institutional autonomy. This has determined the development of significant differences in offerings and the emergence of elite universities that can attract highly-talented faculty and students on the international market, as well as abundant public and private financing, leading to notable socio-economic benefits in their territories. While still suffering from many broad approximations, international rankings do convey the undoubted existence of the phenomenon: the best respected rankings invariably report at least 18 universities from the UK and the USA among the top 20 positions. This is the result of the processes of selection that competitive systems have developed over the span of decades. Other countries show a much different situation: historic choices for greater levels of state regulation have in many cases resulted in the absence or insufficiency of incentive systems, and failed to encourage any qualitative differentiation in what universities offer. This has had important negative consequences in the areas of support to industrial competitiveness, and for the social and economic development of the nation itself.

Particularly in Italy, differentiation among universities has been strongly opposed. In the name of equity and guarantees of equal opportunity for all, legislators designed and nurtured a system of research and higher education while they were unaware of the challenges that the globalized world held in store. The result is a pattern of highly concentrated distribution of research performance within the individual universities, similar to the pattern for national research population as whole. Accompanying this, the variability of performance among universities is less than the variability of performance for individual professors within each university. In synthesis, with the exception of the tiny Scuola Superiore advanced studies institutes, and a very few tiny private universities, Italy has no elite institutions, much less any world-class universities. No international university ranking lists an Italian university above 150$^{th}$ place. The scarce qualitative differentiation in educational offerings and the absence of elite universities results in weaker impact on the competiveness of the productive system, hampers economic development, and has also clearly limited national occupational and economic mobility, which is among the lowest for wealthy countries. University degrees retain a legal value, such that without a degree it is impossible to work in a wide number of professions, or to participate in many job competitions in public administration. But this same system, administered together with the wide-spread favoritism[16], actually hampers advancement of talented youth from the "less influential" classes, since they cannot obtain degrees from elite universities to demonstrate their capacities and value against equally titled but "backed" graduates.

It is widely understood that in the knowledge-based economy, a strong higher education system translates into a decisive contribution to competitiveness of the national system, economic growth and social mobility. Given the current dramatic economic juncture, and emergence of global competition, we ask what can be done to make up the accumulated gap. To unlock the reform process, the most important driving force for modernizing higher education in Italy and similar countries could be a favorable regulatory framework that fosters competition. The reforms initiated by the Italian government in this sense are still very timid. Periodic national assessment exercises were introduced in 2006: the initiative is admirable, but these still show fundamental flaws that limit their effectiveness. First, the resources allocated to universities on the basis of results are still too low to stimulate competition and improvement; second, the resources are allocated in function of average merit of the universities and not to the individual professors. If the policy-maker's intent is to optimize the socio-economic impact of public research financing through allocating more resources to universities with higher average efficiency, but the dispersion of performance within universities is still high (as in Italy), then there is the risk that the macroeconomic objective will never be met, unless the universities in turn carry out internal allocation of resources on performance principles. In fact, these principles are not followed in Italy, because: i) national assessment exercises do not measure individual performance; ii) the majority of professors (we recall that 77% of them produce only 23% of scientific advancement) are hostile to change in this direction. The risk of such failure is much less in systems with strong differentiation among universities, and where the variability of performance within institutions is much less than between them. In a context like Italy's, it would be much more effective to provide individual evaluations and allocate resources directly to single researchers, as occurs in nations that primarily allocate their research funding on the basis of calls for proposals. A further and still more courageous step for Italy would be to link the professors' actual salaries to the quality of their teaching and research. Similar incentives have recently been introduced in the public administration sector, but not yet in higher education, which is itself an indication of the reigning conservative forces. There is room for further and much needed initiatives, which would be greatly effective in stimulating greater competition and continuous improvement: liberation of resources cutting unproductive faculty (apparently about 25% of the total); abolition of the legal value of undergraduate degrees; and abolition of national public competitions for initial hiring and university career advancement.

[16] In the 2011 World Democracy Audit for freedom from corruption, Italy ranked 51$^{st}$ out of 180 countries. http://www.worldaudit.org/corruption.htm, last accessed on January 21, 2013.

The combination of Italy's current financial crisis and the attitude of sacrifice that its citizens are called to support would seem to represent an ideal juncture for the current government to gain public legitimacy for undertaking the necessary structural reforms of the higher education system. However the authors are skeptical that the proper coincidence of conditions and will have arrived, or that if attempts were made, there would be results within acceptable times. Even if very substantial reforms are implemented, it would then take decades for the growth of elite universities that could stop the brain drain, and reverse it attracting foreign talent and resources. While we sincerely hope for the suggested reforms, we suggest an intervention that could proceed independently, and would still be complementary to any other measures taken: an intervention for the birth of spin-off universities, newly staffed by migration of only the top scientists from the existing universities and national research system. This intervention would be much less unpopular because it would not touch vested interests, and so would encounter less resistance to implementation. The asset of the proposal is its effectiveness in rapidly achieving what competitive systems have produced over the span of decades. The necessary policy should favor balanced regional distribution of such top universities and possibly their specialization based on the characteristics of their regions. Our simulation for the city of Rome has shown the higher quality level of such a spin-off university, budded from three existing ones, compared to all other Italian higher education institutions. This qualitative level could be raised still further by drawing faculty from other universities and research institutions in the area. The top universities thus created would by nature be highly resistant to the favoritism virus and much more inclined to adopt practical and principled strategies, typical of those in competitive systems. In the short term, the high qualitative level of faculty would be able to catalyze the other determinants of an elite university: highly talented students, abundant resources and visionary governance. The costs of such an operation are minimal, since a spin-off university could use the infrastructure of one of the parent universities, freed by the departure of the professors who lack qualifications towards the other parent universities. Given the socio-economic benefits, such an initiative would receive the support of all local stakeholders, including government, the productive system, and the local community.

| UDA | SDS* | Faculty | UDA | SDS* | Faculty |
|---|---|---|---|---|---|
| Mathematics | INF/01 | 1 | Engineering | ICAR/01 | 5 |
| | MAT/02 | 5 | | ICAR/02 | 5 |
| | MAT/03 | 9 | | ICAR/03 | 1 |
| | MAT/05 | 12 | | ICAR/04 | 4 |
| | MAT/06 | 2 | | ICAR/05 | 4 |
| | MAT/07 | 5 | | ICAR/07 | 2 |
| | MAT/08 | 1 | | ICAR/08 | 4 |
| | MAT/09 | 2 | | ICAR/09 | 2 |
| Physics | FIS/01 | 14 | | ICAR/20 | 1 |
| | FIS/02 | 7 | | ING-IND/04 | 2 |
| | FIS/03 | 10 | | ING-IND/06 | 1 |
| | FIS/04 | 3 | | ING-IND/08 | 5 |
| | FIS/05 | 4 | | ING-IND/09 | 1 |
| | FIS/06 | 3 | | ING-IND/11 | 3 |
| | FIS/07 | 3 | | ING-IND/12 | 1 |
| Chemistry | CHIM/01 | 7 | | ING-IND/14 | 2 |
| | CHIM/02 | 9 | | ING-IND/17 | 1 |
| | CHIM/03 | 9 | | ING-IND/22 | 2 |
| | CHIM/04 | 3 | | ING-IND/25 | 1 |
| | CHIM/06 | 7 | | ING-IND/26 | 1 |
| | CHIM/07 | 2 | | ING-IND/28 | 2 |
| | CHIM/12 | 1 | | ING-IND/31 | 3 |
| | SECS-P/13 | 1 | | ING-IND/32 | 3 |
| Earth science | GEO/01 | 2 | | ING-IND/35 | 1 |
| | GEO/02 | 2 | | ING-INF/01 | 9 |
| | GEO/03 | 6 | | ING-INF/02 | 5 |
| | GEO/04 | 1 | | ING-INF/03 | 6 |
| | GEO/05 | 2 | | ING-INF/04 | 6 |
| | GEO/06 | 3 | | ING-INF/05 | 14 |
| | GEO/07 | 1 | | ING-INF/06 | 3 |
| | GEO/08 | 6 | | ING-INF/07 | 2 |
| | GEO/09 | 1 | | SECS-P/02 | 1 |

***Table 1: Distribution of RSU faculty by SDS and UDA***

** SDS codes are accessible at www.disp.uniroma2.it/laboratorioRTT/TESTI/Indicators/ssd1.htm*

| UDA | Full professors | Associate professors | Assistant professors |
|---|---|---|---|
| Mathematics | 10 | 11 | 6 |
| Physics | 18 | 16 | 8 |
| Earth science | 6 | 8 | 9 |
| Chemistry | 11 | 13 | 13 |
| Engineering | 42 | 40 | 36 |
| Total | 87 (35.2%) | 88 (35.6%) | 72 (29.1%) |
| Total Italy | 7175 (33.6%) | 7155 (33.5%) | 7013 (32.9%) |
| Diff. | +1.6% | +2.1% | -3.8% |

***Table 2: Composition of RSU faculty per academic rank and UDA***

| UDA/Univ. | Sapienza | Tor Vergata | Rome Tre | RSU |
|---|---|---|---|---|
| Mathematics | 11 | 11 | 5 | 27 |
| Physics | 25 | 7 | 10 | 42 |
| Earth science | 14 | - | 9 | 23 |
| Chemistry | 27 | 10 | - | 37 |
| Engineering | 58 | 36 | 24 | 118 |
| Total | 135 | 64 | 48 | 247 |

***Table 3: Parent universities for RSU faculty, by UDA***

| UDA/Univ | "Sapienza" | | "Tor Vergata" | | "TRE" | |
|---|---|---|---|---|---|---|
| Mathematics | 40.7% | (0.90) | 40.7% | (1.02) | 18.5% | (1.27) |
| Physics | 59.5% | (1.23) | 16.7% | (0.50) | 23.8% | (1.31) |
| Earth science | 60.9% | (0.85) | 0.0% | (0.00) | 39.1% | (1.40) |
| Chemistry | 73.0% | (1.00) | 27.0% | (0.99) | 0.0% | (0.00) |
| Engineering | 49.2% | (0.85) | 30.5% | (1.06) | 20.3% | (1.50) |
| Total | 54.7% | (0.96) | 25.9% | (0.89) | 19.4% | (1.39) |

***Table 4: Contribution to RSU faculty from the three parent Rome universities (in parentheses the concentration with respect to total faculty)***

| School/University | Average percentile rank | Faculty above $80^{th}$ percentile | Faculty above $90^{th}$ percentile |
|---|---|---|---|
| Mathematics | 88.9 | 89.2 | 59.5 |
| Physics | 91.2 | 90.5 | 71.4 |
| Earth science | 85.6 | 75.4 | 44.9 |
| Chemistry | 95.0 | 100.0 | 77.8 |
| Engineering | 92.9 | 91.3 | 69.6 |
| Total RSU | 88.8 | 84.2 | 57.5 |
| Total “Sapienza” | 40.9 | 14.9 | 7.1 |
| Total “Tor Vergata” | 48.8 | 20.9 | 10.8 |
| Total “TRE” | 47.2 | 21.8 | 10.9 |
| Top public university | 60.7 | 36.5 | 22.8 |
| Top Scuola Superiore | 68.7 | 54.5 | 31.8 |
| Top private university | 69.6 | 44.2 | 29.9 |

***Table 5: Distribution of research performance for the RSU faculty and comparison with the three parent Rome universities and the top national universities***

| FSAc percentile | RSU | Average parent universities | Average Italian universities |
|---|---|---|---|
| =100 | 6.9% (17 of 247) | 0.8% | 1.1% |
| ≥99 | 9.7% (24 of 247) | 1.3% | 1.6% |
| ≥98 | 16.2% (40 of 247) | 2.3% | 2.4% |
| ≥95 | 38.1% (94 of 247) | 5.5% | 5.3% |
| ≥90 | 57.5% (142 of 247) | 9.8% | 11.9% |
| ≥80 | 84.2% (208 of 247) | 19.4% | 21.0% |
| ≥50 | 98.8% (244 of 247) | 49.6% | 47.8% |

***Table 6: Distribution of research performance for RSU faculty, by percentile class, and comparison with the three Rome parent universities and all Italian universities***

| Academic rank | Yearly average cost (k€) | Equivalent cost |
|---|---|---|
| Full professor (confirmed) | 124.939 | 2.783 |
| Full professor (probationary) | 94.442 | 2.103 |
| Associate professor (confirmed) | 90.622 | 2.018 |
| Associate professor (probationary) | 68.469 | 1.525 |
| Assistant professor (confirmed) | 68.844 | 1.533 |
| Assistant professor (probationary) | 44.899 | 1 |

***Table 7: Yearly cost of Italian professors, by academic rank (average values 2004-2008)***

| SDS | UDA | P* | Rank | Percent. | SDS | UDA | P* | Rank | Percent. |
|---|---|---|---|---|---|---|---|---|---|
| MAT/02 | | 0.39 | 2 out of 40 | 97.4 | ICAR/01 | | 0.11 | 8 out of 35 | 79.4 |
| MAT/03 | | 0.39 | 1 out of 50 | 100 | ICAR/02 | | 0.15 | 9 out of 39 | 78.9 |
| MAT/05 | | 1.11 | 1 out of 54 | 100 | ICAR/03 | | 0.21 | 4 out of 31 | 90 |
| MAT/06 | Mathematics | 0.57 | 4 out of 39 | 92.1 | ICAR/04 | | 0.15 | 1 out of 22 | 100 |
| MAT/07 | | 1.12 | 1 out of 49 | 100 | ICAR/05 | | 0.06 | 6 out of 27 | 80.8 |
| MAT/08 | | 0.58 | 4 out of 49 | 93.8 | ICAR/07 | | 0.06 | 5 out of 39 | 89.5 |
| MAT/09 | | 0.36 | 9 out of 36 | 77.1 | ICAR/08 | | 0.78 | 1 out of 44 | 100 |
| INF/01 | | 1.42 | 1 out of 52 | 100 | ICAR/09 | | 0.19 | 6 out of 42 | 87.8 |
| FIS/01 | | 0.99 | 1 out of 54 | 100 | ICAR/20 | | 0.01 | 9 out of 18 | 52.9 |
| FIS/02 | | 2.05 | 2 out of 39 | 97.4 | ING-IND/04 | | 0.17 | 2 out of 11 | 90 |
| FIS/03 | | 2.13 | 1 out of 43 | 100 | ING-IND/06 | | 0.86 | 1 out of 18 | 100 |
| FIS/04 | Physics | 0.44 | 2 out of 32 | 96.8 | ING-IND/08 | | 0.38 | 2 out of 35 | 97.1 |
| FIS/05 | | 1.52 | 2 out of 26 | 96 | ING-IND/09 | | 0.52 | 1 out of 29 | 100 |
| FIS/06 | | 0.15 | 9 out of 24 | 65.2 | ING-IND/11 | | 0.20 | 2 out of 41 | 97.5 |
| FIS/07 | | 0.67 | 1 out of 46 | 100 | ING-IND/12 | | 0.09 | 7 out of 21 | 70 |
| CHIM/01 | | 1.52 | 1 out of 44 | 100 | ING-IND/14 | Engineering | 0.16 | 15 out of 36 | 60 |
| CHIM/02 | | 1.58 | 3 out of 43 | 95.2 | ING-IND/17 | | 0.40 | 1 out of 33 | 100 |
| CHIM/03 | | 0.93 | 5 out of 47 | 91.3 | ING-IND/22 | | 1.64 | 1 out of 40 | 100 |
| CHIM/04 | Chemistry | 0.43 | 6 out of 24 | 78.3 | ING-IND/25 | | 0.80 | 1 out of 28 | 100 |
| CHIM/06 | | 0.99 | 2 out of 50 | 98.0 | ING-IND/26 | | 0.84 | 2 out of 16 | 93.3 |
| CHIM/07 | | 1.28 | 3 out of 38 | 94.6 | ING-IND/28 | | 0.03 | 2 out of 7 | 83.3 |
| CHIM/12 | | 0.55 | 2 out of 30 | 96.6 | ING-IND/31 | | 0.27 | 13 out of 37 | 66.7 |
| SECS-P/13 | | 1.06 | 1 out of 17 | 100 | ING-IND/32 | | 0.10 | 9 out of 25 | 66.7 |
| GEO/01 | | 0.61 | 2 out of 31 | 96.7 | ING-IND/35 | | 0.20 | 3 out of 31 | 93.3 |
| GEO/02 | | 0.17 | 7 out of 40 | 84.6 | ING-INF/01 | | 1.46 | 1 out of 43 | 100 |
| GEO/03 | | 0.96 | 1 out of 32 | 100 | ING-INF/02 | | 1.07 | 2 out of 37 | 97.2 |
| GEO/04 | | 0.13 | 7 out of 36 | 82.9 | ING-INF/03 | | 0.93 | 2 out of 42 | 97.6 |
| GEO/05 | Earth science | 0.28 | 1 out of 41 | 100 | ING-INF/04 | | 0.83 | 3 out of 44 | 95.3 |
| GEO/06 | | 0.51 | 2 out of 29 | 96.4 | ING-INF/05 | | 0.60 | 4 out of 59 | 94.8 |
| GEO/07 | | 1.24 | 1 out of 32 | 100 | ING-INF/06 | | 0.18 | 12 out of 23 | 50 |
| GEO/08 | | 0.50 | 3 out of 30 | 93.1 | ING-INF/07 | | 0.14 | 20 out of 36 | 45.7 |
| GEO/09 | | 0.41 | 2 out of 29 | 96.4 | SECS-P/02 | | 1.20 | 1 out of 49 | 100 |

***Table 8: Absolute value and national rank (absolute and percentile) for productivity of the RSU SDSs***

| UDA | $P_U$ | rank | percentile |
|---|---|---|---|
| Mathematics | 4.23 | 2 out of 65 | 98.4 |
| Physics | 3.55 | 1 out of 62 | 100 |
| Chemistry | 2.73 | 3 out of 60 | 96.6 |
| Earth science | 3.31 | 1 out of 49 | 100 |
| Engineering | 5.30 | 2 out of 69 | 98.5 |

***Table 9: Absolute value and national rank (absolute and percentile) of productivity for RSU UDAs***